\documentclass[]{spie}  %>>> use for US letter paper
\usepackage[]{graphicx}

\title{Measuring the flatness of focal plane for very large mosaic CCD camera} 

\author{Jiangang Hao\supit{*}, Juan Estrada\supit{$\dagger$}, Herman Cease, H. Thomas Diehl, Brenna L. Flaugher, Donna Kubik, Keivin Kuk, Nickolai Kuropatkine, Huan Lin, Jorge Montes, Vic Scarpine, Ken Schultz and William Wester for the DES collaboration
\skiplinehalf
Center for Particle Astrophysics, Fermi National Accelerator Laboratory, P.O. Box 500, Batavia, IL 60510, USA
}

\authorinfo{\supit{*}jghao@fnal.gov; \supit{$\dagger$}estrada@fnal.gov\\FERMILAB-CONF-10-191-A}
\begin{document} 
  \maketitle 

%%%%%%%%%%%%%%%%%%%%%%%%%%%%%%%%%%%%%%%%%%%%%%%%%%%%%%%%%%%%% 
\begin{abstract}

Large mosaic multiCCD camera is the key instrument for modern digital sky survey. DECam is an extremely red sensitive 520 Megapixel camera designed for the incoming Dark Energy Survey (DES). It is consist of sixty two 4k$\times$2k and twelve 2k$\times$2k 250-micron thick fully-depleted CCDs, with a focal plane of 44 cm in diameter and a field of view of 2.2 square degree. It will be attached to the Blanco 4-meter telescope at CTIO. The DES will cover 5000 square-degrees of the southern galactic cap in 5 color bands (g, r, i, z, Y) in 5 years starting from 2011.

To achieve the science goal of constraining the Dark Energy evolution, stringent requirements are laid down for the design of DECam. Among them, the flatness of the focal plane needs to be controlled within a 60-micron envelope in order to achieve the specified PSF variation limit. It is very challenging to measure the flatness of the focal plane to such precision when it is placed in a high vacuum dewar at 173 K.  We developed two image based techniques to measure the flatness of the focal plane. By imaging a regular grid of dots on the focal plane, the CCD offset along the optical axis is converted to the variation the grid spacings at different positions on the focal plane. After extracting the patterns and comparing the change in spacings, we can measure the flatness to high precision. In method 1, the regular dots are kept in high sub micron precision and cover the whole focal plane. In method 2, no high precision for the grid is required. Instead, we use a precise XY stage moves the pattern across the whole focal plane and comparing the variations of the spacing when it is imaged by different CCDs. Simulation and real measurements show that the two methods work very well for our purpose, and are in good agreement with the direct optical measurements.

\end{abstract}

%>>>> Include a list of keywords after the abstract 


%%%%%%%%%%%%%%%%%%%%%%%%%%%%%%%%%%%%%%%%%%%%%%%%%%%%%%%%%%%%%
\section{INTRODUCTION}
\label{sec:intro}  % \label{} allows reference to this section

The discovery of cosmic acceleration from type Ia supernovae experiments~\cite{riess98,perlmutter98} in late 1990s has pushed cosmology in to a new era. Independent experiments from Cosmic Microwave Background Radiation measured by WMAP~\cite{spergel03}, large scale galaxy clustering ~\cite{tegmark04} and Baryon Acoustic Oscillation (BAO)~\cite{eisenstein05} from Sloan Digital Sky Survey (SDSS) strongly suggest that there exists a new substance, dark energy, which has negative pressure and is about two thirds of the total energy density in the Universe. Dark Energy Survey (DES)~\cite{des} is a project dedicated to understanding the nature of Dark Energy by imaging 5000 square degree southern galactic cap in 5 color bands (g, r, i, z, Y) and up to redshift 1.3. The data will allow us to extract cosmological information on the dark energy from 1) cluster counting and spatial distribution of clusters at redshift from 0.1 to 1.3; 2) Baryon acoustic oscillation; 3) weak lensing measurements on several redshift shells to redshift $\sim$ 1; and 4) 2000 supernovae at redshift from 0.3 to 0.8. These joint cosmological probes will constrain the equation of state of dark energy to 5 - 15\% and its evolution to 30\% ({\tt http://www.darkenergysurvey.org}), greatly improving our understanding of dark energy.

The key instrument for DES is the DECam, a red sensitive 520 Megapixel camera consisting of sixty two 4k$\times$2k and twelve 2k$\times$2k 250-micron thick fully-depleted CCDs. The focal plane of the camera is 44 cm in diameter and spans a field of view of 2.2 square degree. DECam is assembled, calibrated and tested at Fermi National Accelerator Laboratory and will be shipped to Cerro Tololo Inter-American Observatory (CTIO) in Chile, where it will be mounted to the Blanco 4-meter telescope in 2011. 

The large focal plane of DECam consists of 74 CCD detectors. If some CCDs offset from the focal plane along the optical axis, they will be out of the best focus, leading to "blurring" objects on the image. Such blurring will degrade the precision of galaxy shape measurements, which is crucial for tight constraints on dark energy parameters using gravitational lensing technique. As a results, DES requires all CCDs to be within a 60-micron envelope from the focal plane. In the following, we will refer this as focal plane flatness requirement. 

Measuring the flatness of the focal plane turns out to be a very challenging task. The reason is that the focal plane (with CCD mounted) is kept inside a dewar (the camera chamber) at low temperature (173 K) and high vacuum. There is a glass window that insulate the focal plane inside the camera chamber. Since we cannot touch the CCDs mechanically, we will have to use optics based methods. For example, the Micro-Epsilon Opto-NCDT 2400 has been used to measure the flatness of focal plane~\cite{derlo06,diehl08}. However, the glass window as well as the CCD surface reflection properties may introduce unknown uncertainties into the measurement. In this paper, we present two image based measurements, which directly measure the variation of a grid of dots at different positions on the focal plane. This variations can be converted to the offset of CCDs along the optical axis. Moreover, the purpose of flatness requirement is to ensure the image does not blur due to the CCD offset from focal plane, therefore the image based method directly measure the degree of "blurring". These two methods also provide an independent cross check to the measurements from Opto-NCDT. 

\section{Measuring the flatness of focal plane based on image}

The basic idea for image-based flatness measurement is to image a uniform grid of dots using the CCDs and then check the variations of the separations between the dots. If some detectors are offset from the focal plane, the separations between the dots imaged by them will increase or decrease, depending on which way they offset from the focal plane. The most important thing for this kind of measurements is to image a uniform grid of dots and extract their positions precisely. Meanwhile, to avoid the deformation due to lens system, we should not introduce any lens for imaging unless we have calibrated them very well. Ideally, if we can image a grid of dots on the whole focal plane and have enough dots on each CCD, we will simply need to analyse the change of separations and convert them to the offset of CCDs from focal plane. In practice, this requires the precision of the grid to be at sub-micron level for our purpose, which is very challenging. We therefore develop an alternative method that does not require a large uniform grid of dots to very high precision. Instead, we use a precision XY stage to move the pattern across the focal plane and image it with different CCDs. We then cross match the patterns and compare the change of the same separations imaged by different CCDs and at different positions. In the following, we will first demonstrate how to measure the offset using a large uniform grid and then we show the technique using XY stage. For the first method, we have not yet perform any actual measurements and will only show some simulation study. For the second method, we present some real measurement results. 

\subsection{Simulation study}

To show the first method really works in a realistic setting, we simulate a large regular grid of dots on the full focal plane and then run our data processing pipeline to extract the positions. Among the 74 CCDs, 64 of them are used for imaging the sky and the rest are used for focusing and guiding. In the simulation, we assume a uniform grid of dots is imaged by the CCDs through a pinhole. The parameters of the simulation are specified as following:
\begin{itemize}
\item The distance between the pinhole and the focal plane is 150 mm and the distance between source and pinhole is 14.286 mm. This leads to a magnification of 10.5.
\item Grid position precision on the image is 1 micron, and the corresponding precision of the grid on the source plane is $\sim$ 0.1 micron.
\item CCD pixel size is 15 microns.
\item Separation between the neighbouring dots is 4.2 mm $\sim$ 280 pixels.
\item The DES specification for the focal plane non-flatness is 60 microns (4 pixels). With our parameter setup, this offset will produce an increase/decrease of 0.106 pixel for the separations of neighbouring dots and $0.106 \times 8 = 0.804$ pixels for separations between every 8 dots.
\item The CCD 21, 22 and 28 are set backward by 4 pixels while CCD 35, 41 and 42 are set forward by 4 pixels.
\end{itemize}

\begin{figure}
   \begin{center}
   \begin{tabular}{r}
   \includegraphics[height=10cm]{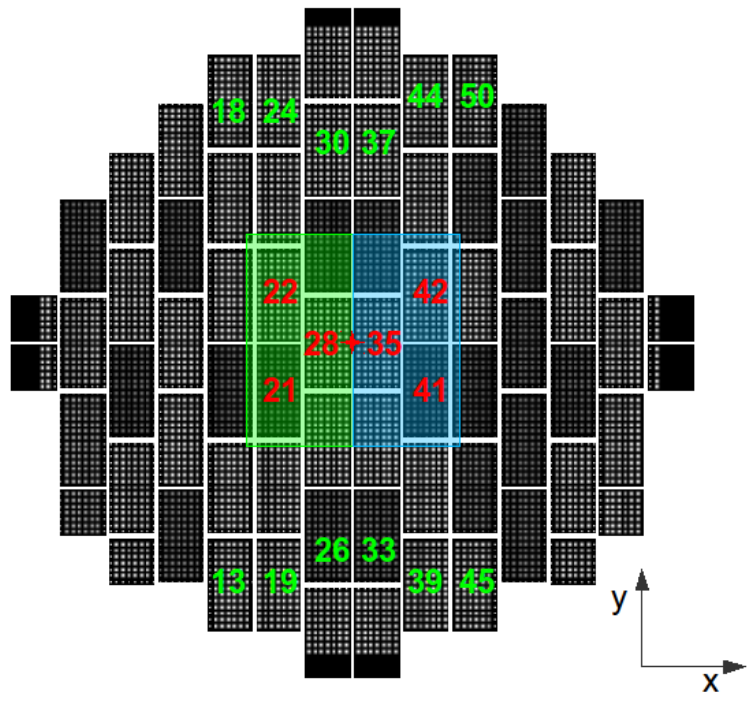}
   \end{tabular}
   \end{center}
   \caption[projector] 
%>>>> use \label inside caption to get Fig. number with \ref{}
   { \label{fig:simfp} 
A uniform grid of dots are generated on the multiCCD. }
   \end{figure}

\begin{figure}
   \begin{center}
   \begin{tabular}{r}
   \includegraphics[width=12cm,height=7cm]{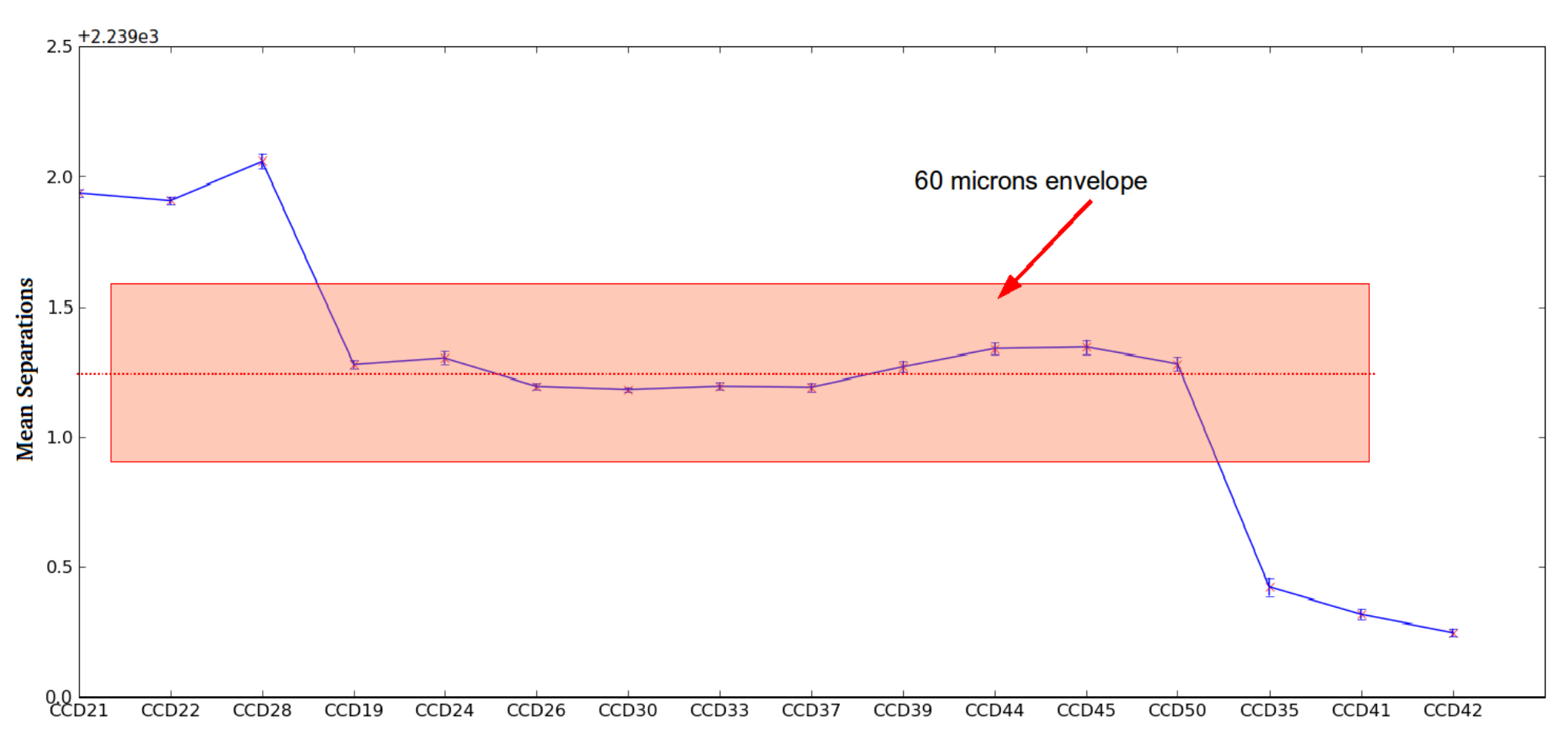}
   \end{tabular}
   \end{center}
   \caption[projector] 
%>>>> use \label inside caption to get Fig. number with \ref{}
   { \label{fig:simsp} 
Identify the CCD offset by looking at the mean variations of spacing between every 8 dots on each CCD. CCD 21, 22 and 28 are offset backward 4 pixels from the focal plane and CCD 35, 41 and 42 are offset forward 4 pixels from the focal plane. They are all falling outside of the 60-micron envelope of our specification. But clearly, we can detect the offset even if they are only offset by 2 pixels. }
   \end{figure}

The resulting image is shown in Figure~\ref{fig:simfp}. Next, we run SExtractor~\cite{bertin96} on the image and extract the positions of the simulated dots. Then, we calculate the mean separation between every 8 dots on a number of CCDs including the ones we add offset deliberately in our simulation. In Figure~\ref{fig:simsp}, we present the measurement results. From the plot, one can clearly identify the CCDs being offset from the focal plane. Therefore, we conclude that this method can work very well to measure the flatness if we can keep the precision of the grid to 1 micron level on the image.

%-------------

%-------------

\subsection{Measuring the flatness by moving a fixed pattern using a precision XY stage}

\subsubsection{Theory}
\begin{figure}
   \begin{center}
   \begin{tabular}{c c}
   \includegraphics[height=5cm]{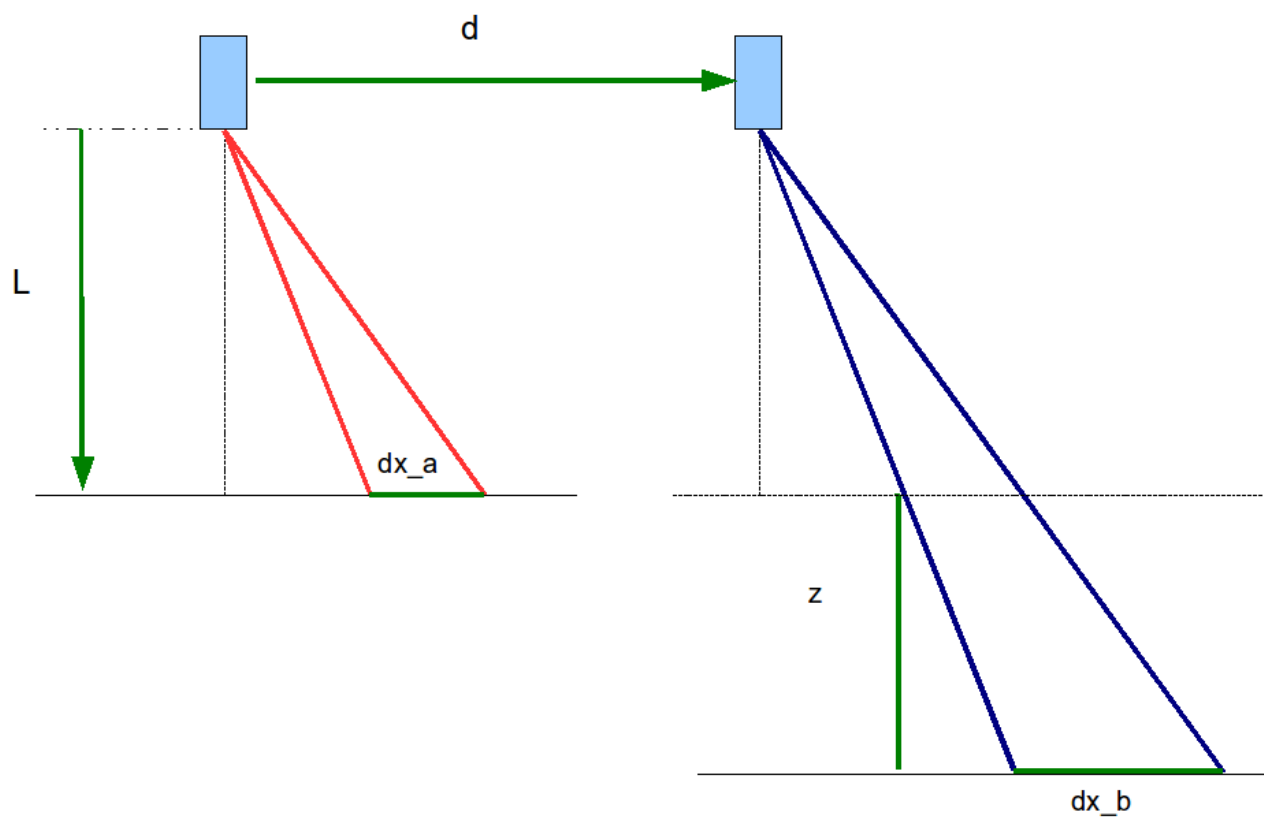}&\includegraphics[height=5cm]{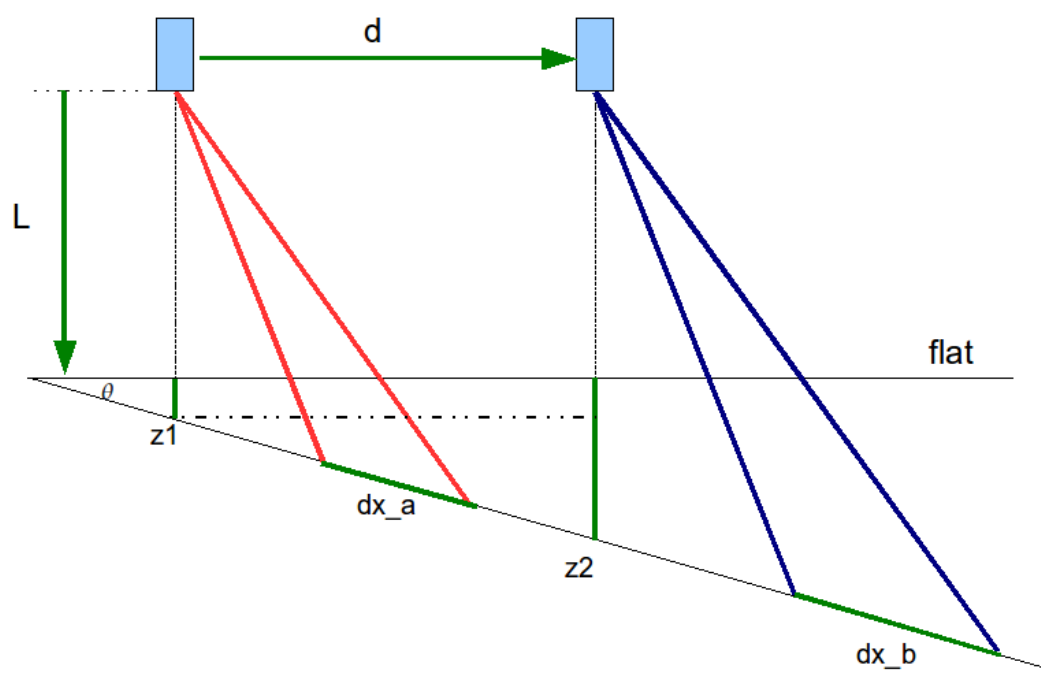}\\
    (a) & (b)
   \end{tabular}
   \end{center}
   \caption[projector] 
%>>>> use \label inside caption to get Fig. number with \ref{}
   { \label{fig:theory} 
(a) Offset between two CCDs when there is no significant tilt on each of them. (b) The tilt of a single CCD.}
   \end{figure}

In this section, we derive the formula that relate the measurable quantities and offset of the CCD from the focal plane. First, we assume that the CCDs are not tilt significantly compared to their offset from the focal plane (see Figure~\ref{fig:theory} (a)). In this case, we have the following formula for the average offset. 

\begin{equation}\label{equation:notilt}
<z>=\frac{L}{N}\sum_i^N \frac{dx_{b}^{(i)}-dx_{a}^{(i)}}{dx_a^{(i)}}
\end{equation}
\noindent where $dx_a^{(i)}$ and $dx_b^{(i)}$ are the length of the segments between the dots. L is the distance between the pinhole and the focal plane. On the other hand, when there is a big tilt for a CCD (see Figure~\ref{fig:theory} (b)), we can also derive the formula that relate the average tangent of the tilt angle $\theta$ to the measurable quantities. The equation is as following: 

\begin{equation}\label{equation:tilt}
<\tan \theta>=\frac{L}{d N}\sum_i^N \frac{dx_{b}^{(i)}-dx_{a}^{(i)}}{dx_b^{(i)}}
\end{equation}

\noindent where d is the distance between the two positions of the pinhole. 

In Equations (\ref{equation:notilt}) and (\ref{equation:tilt}), the right hand side are all measurable quantities. We will measure these quantities based on images.

\subsubsection{Instruments and setup}

\begin{figure}
   \begin{center}
   \begin{tabular}{c}
   \includegraphics[height=4cm]{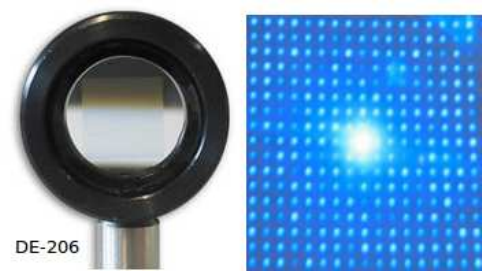}
   \includegraphics[height=3.85cm]{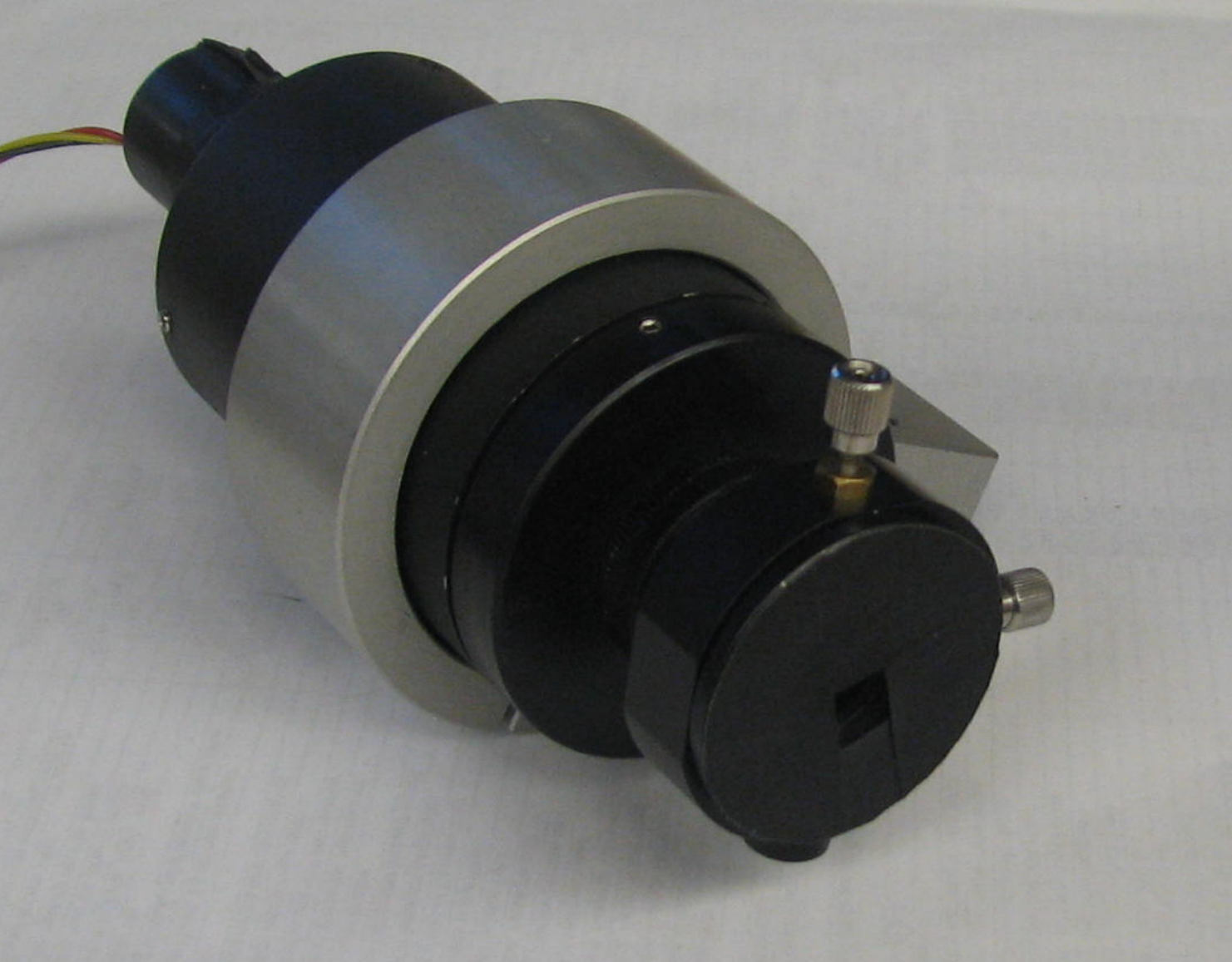}
   \end{tabular}
   \end{center}
   \caption[projector] 
%>>>> use \label inside caption to get Fig. number with \ref{}
   { \label{fig:projector} 
Left: the diffractive grating DE-206; middle: the grid generated by shooting a laser beam on DE-206; right: the final ``star'' projector with the DE-206 and laser encapsulated. We mask out the central bright spot and outer region by placing a baffle in the forefront of the projector.}
   \end{figure}

\begin{figure}
   \begin{center}
   \begin{tabular}{c}
   \includegraphics[width=6cm]{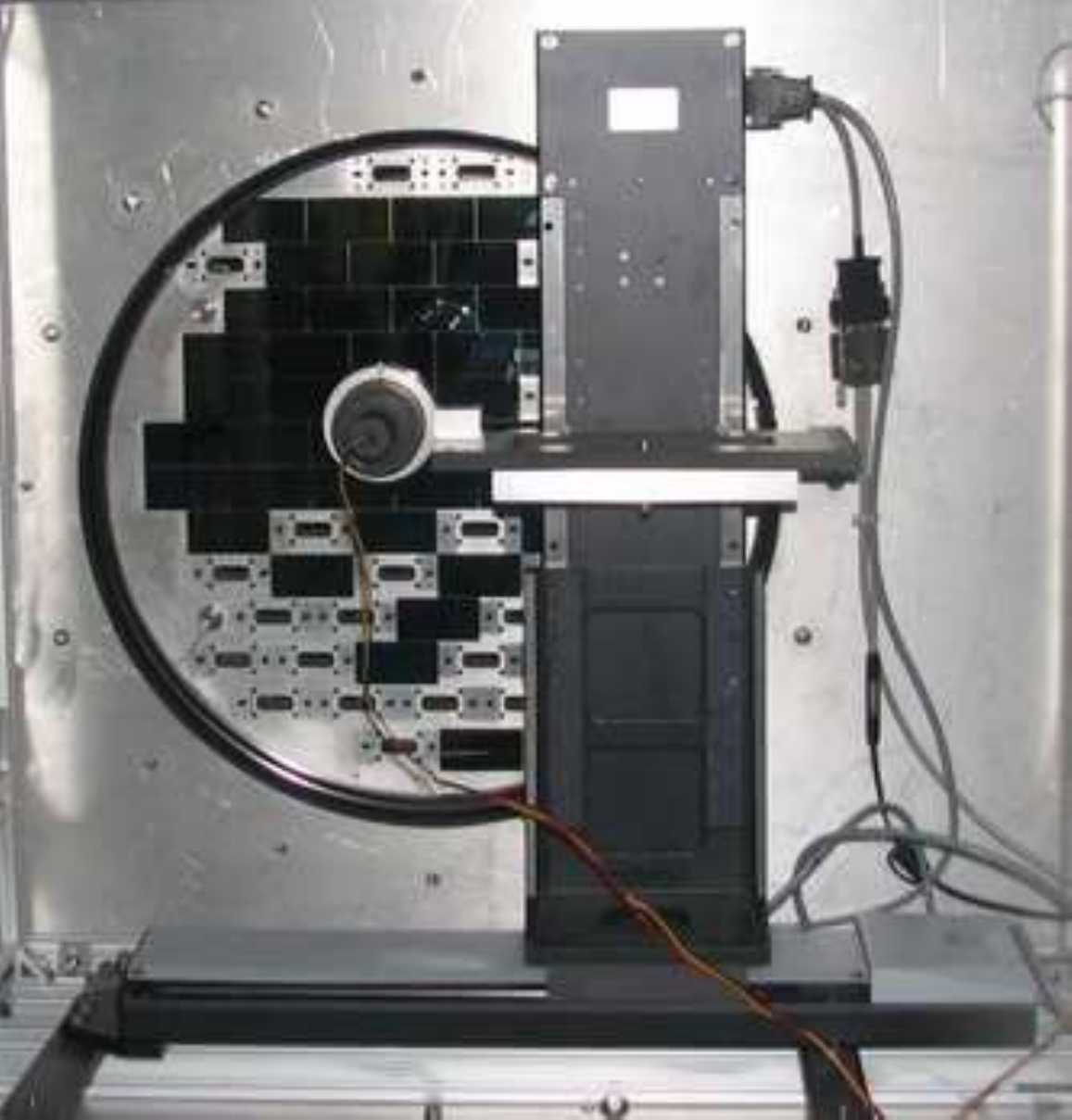}
   \includegraphics[height=7cm]{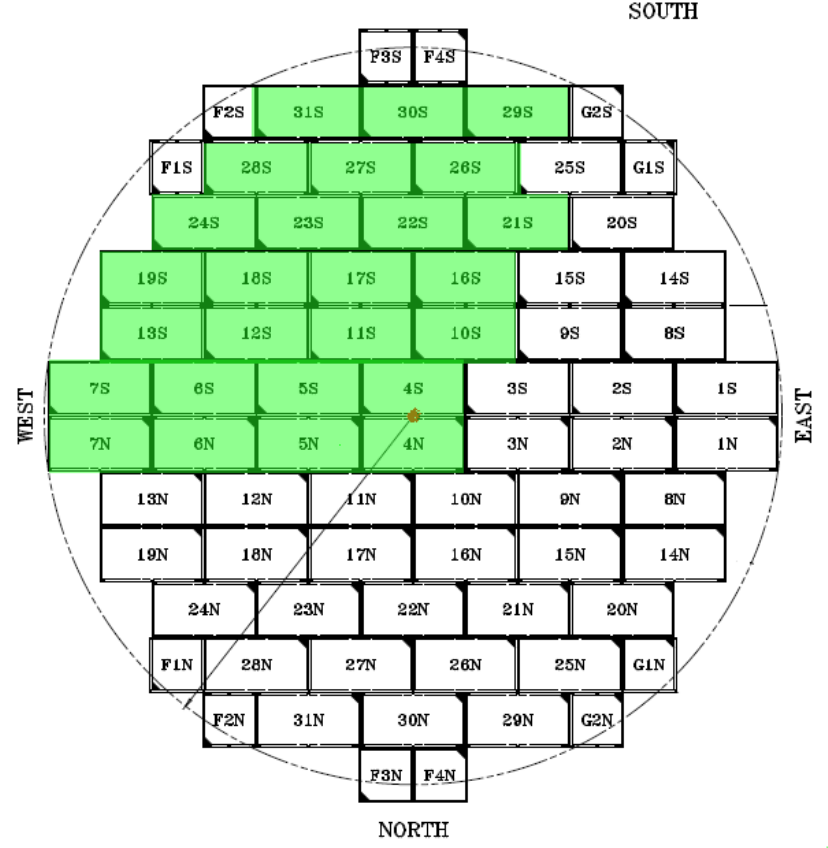}
   \end{tabular}
   \end{center}
   \caption[xystage] 
%>>>> use \label inside caption to get Fig. number with \ref{}
   { \label{fig:xystage} 
Left: The XY-stage with the star projector mounted in the front of the focal plane. We use it to move the projectors across the focal plane and take images using different CCDs. Right: the layout of the DECam focal plane. The region covered by green color are where working CCDs are mounted. In our measurements, we will focus on these CCDs. 
}
   \end{figure}

A key part of this measurement is to generate a grid of dots and image it with CCDs. One simple way to achieve this is to use a pinhole. Another way is to use a diffractive grating to split a laser beam into a grid of dots. In our implementation, we choose the grating to project a grid of dots on the focal plane. The grating we used is DE-206 and is made by HOLOEYE Corporation ({\tt http://www.holoeye.com/index.html}). When shooting a laser beam on it, a grid of uniform dots will be generated. The spot where laser hit the grating is equivalent to the pinhole position of a pinhole camera. So our previous theory can be directly applied to this case too. We encapsulate the DE-206 and a blue laser (480 nm) together and assemble a ``star'' projector. In Figure \ref{fig:projector}, we show ``star'' projector and its components. We mount the ``star'' projector onto a XY stage, which is mounted in the front of the focal plane as shown in Figure~\ref{fig:xystage}.

%-------------

%-------------

\subsubsection{Setup Calibration}

\begin{figure}
   \begin{center}
   \begin{tabular}{c}
   \includegraphics[width=8cm]{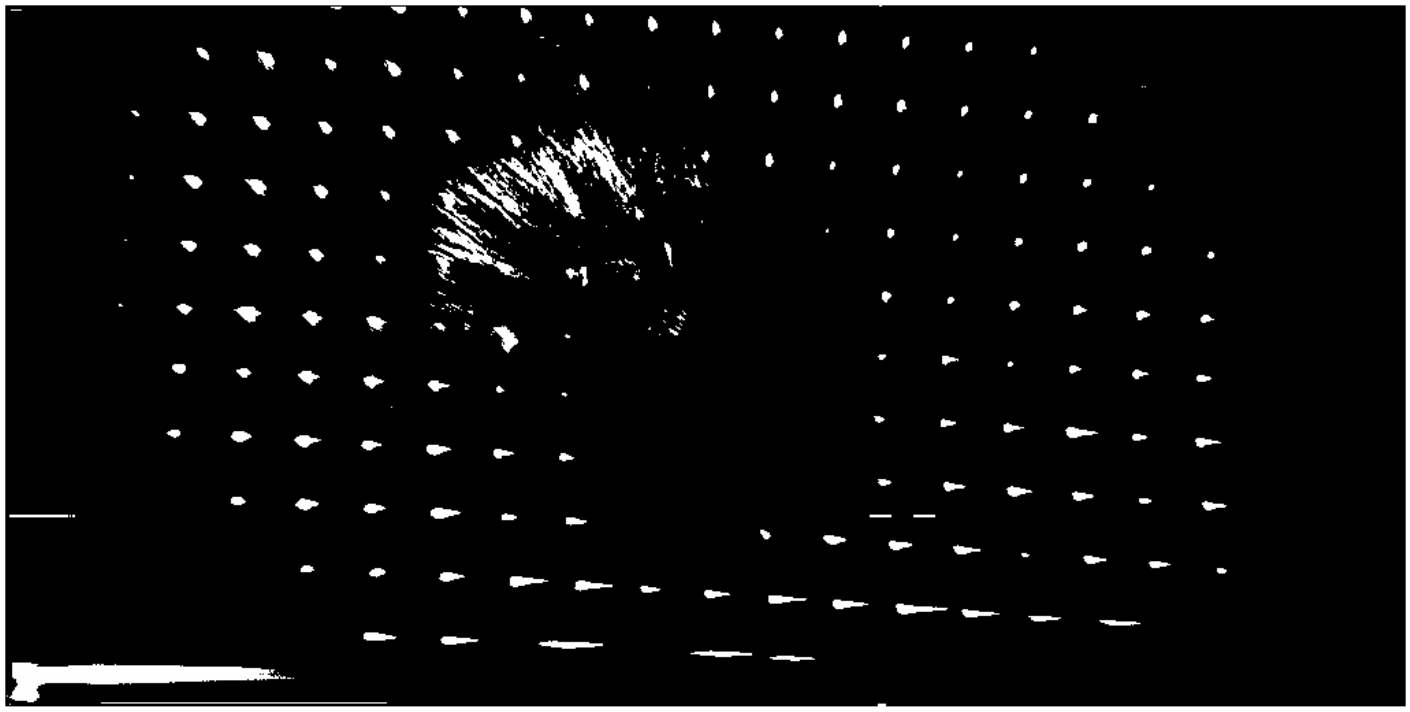}
    \end{tabular}
   \end{center}
   \caption[xystage] 
%>>>> use \label inside caption to get Fig. number with \ref{}
   { \label{fig:star}``Stars'' projected by the ``star'' projector onto one CCD. Note that the ``star'' pattern is produced via diffraction, the center is dominated by the zeroth order diffraction and show very bright spot. In our application, we masked it out but not perfectly, leaving some residuals. In our data analysis, we choose a brightness threshold to remove them. 
}
   \end{figure}

\begin{figure}
   \begin{center}
   \begin{tabular}{c}
   \includegraphics[height=10cm]{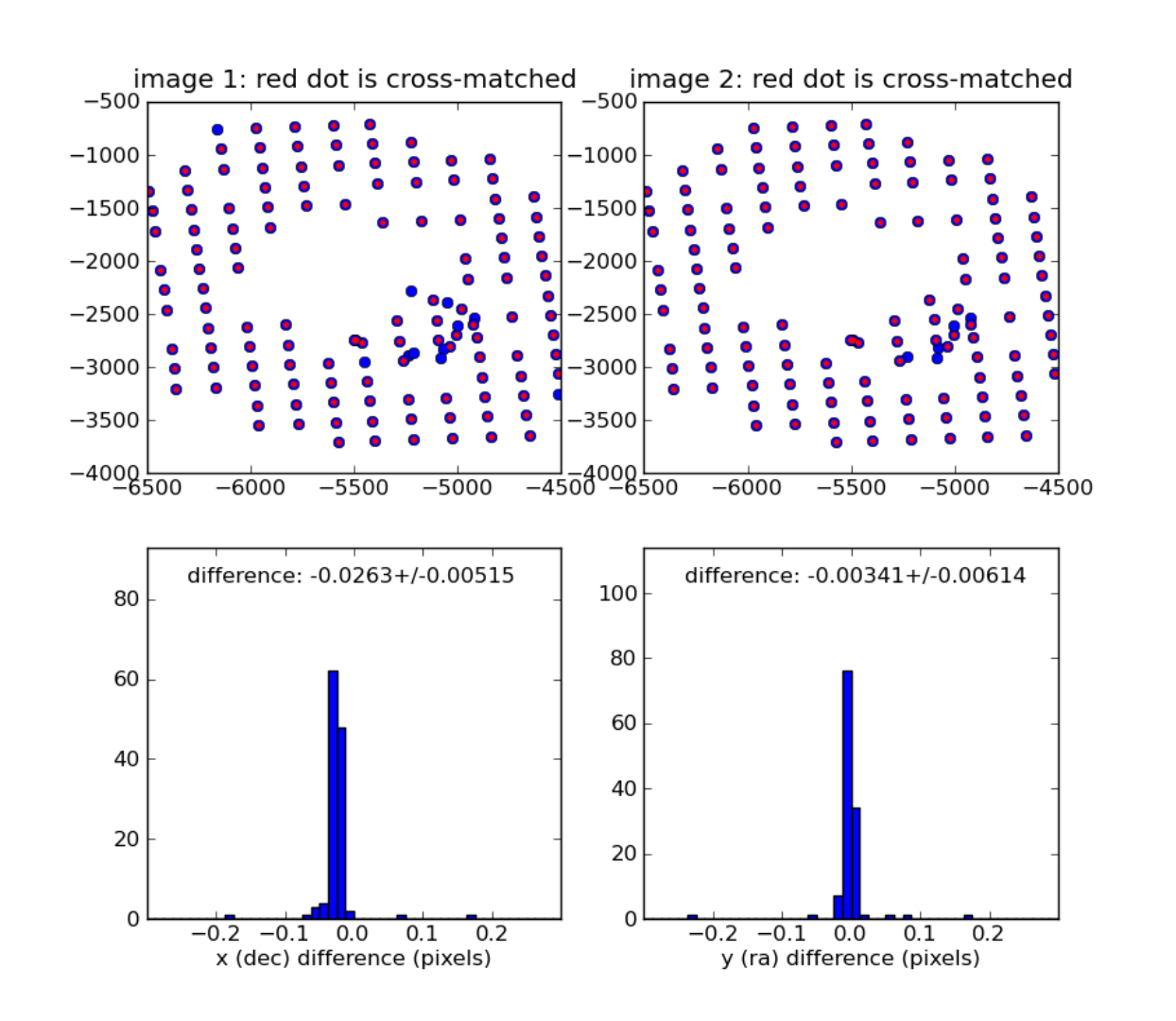}
    \end{tabular}
   \end{center}
   \caption[xystage] 
%>>>> use \label inside caption to get Fig. number with \ref{}
   { \label{fig:repeat} 
Top two panels shows the SExtracted positions of the ``stars'' on the two images. The bottom two panels show the distribution of the position difference in x and y direction.
}
   \end{figure}

Before we start our measurements, we need to make sure that the XY stage can work stably and we can extract the positions of the stars reliably. To do this, we firstly move the ``star'' projector to position 17S (see right panel in Figure~\ref{fig:xystage}) and take an image. The CCD readouts is though the SISPI system~\cite{klaus08}. Then, we move it to somewhere else. Next, we move it back to 17S and take an image. Next, we run SExtractor on the images and build a catalog for the positions of ``stars''. In Figure~\ref{fig:star}, we show a CCD image for the ``stars''. In Figure~\ref{fig:repeat}, we show the difference of the positions of the stars from the two images. They are in very good agreement and thus inform us that the XY stage can work very reliably and our data extraction pipeline works well. 

%-------------

\subsubsection{Measurements}

\begin{figure}
   \begin{center}
   \begin{tabular}{c c}
   \includegraphics[width=7cm,height=5cm]{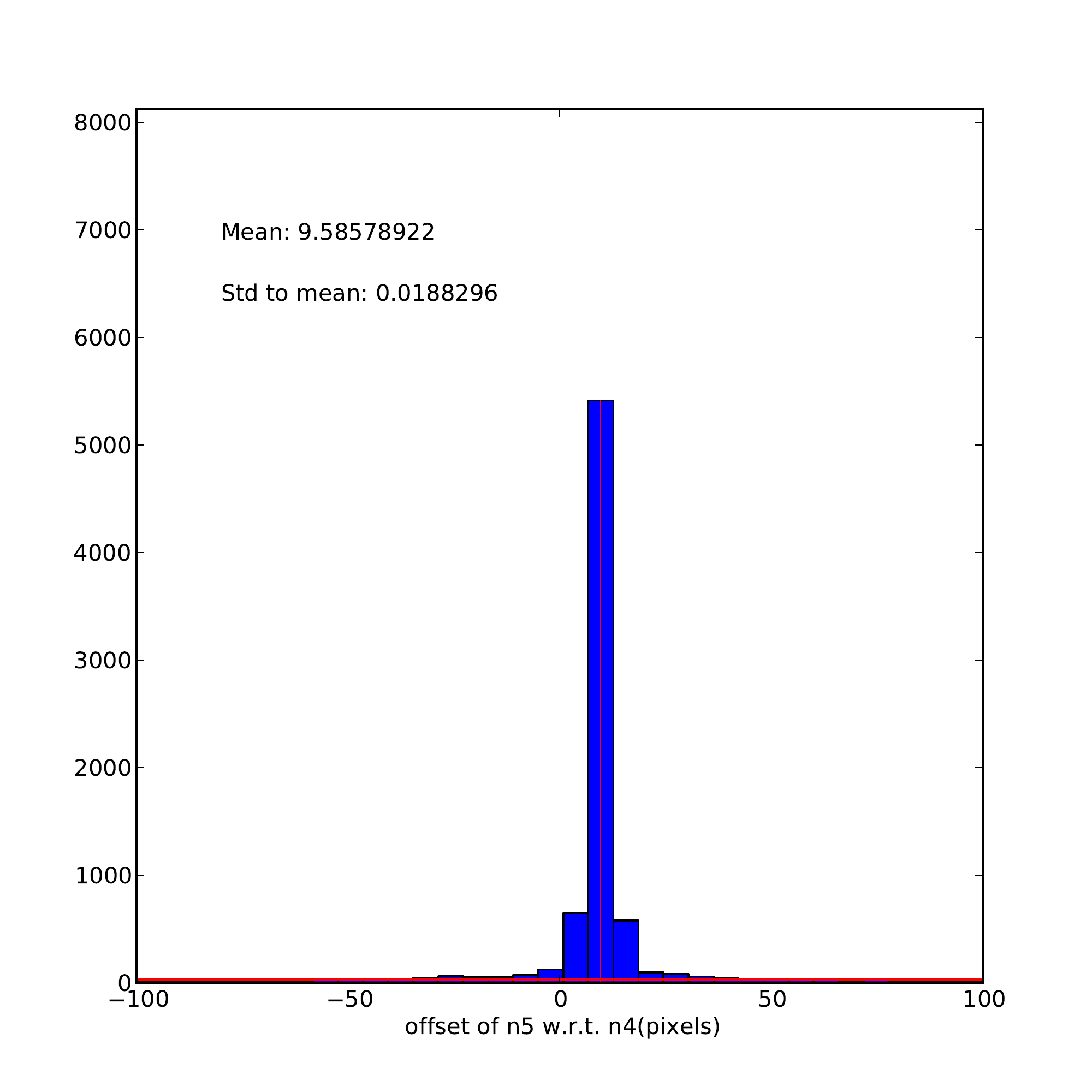} & \includegraphics[width=7cm,height=5cm]{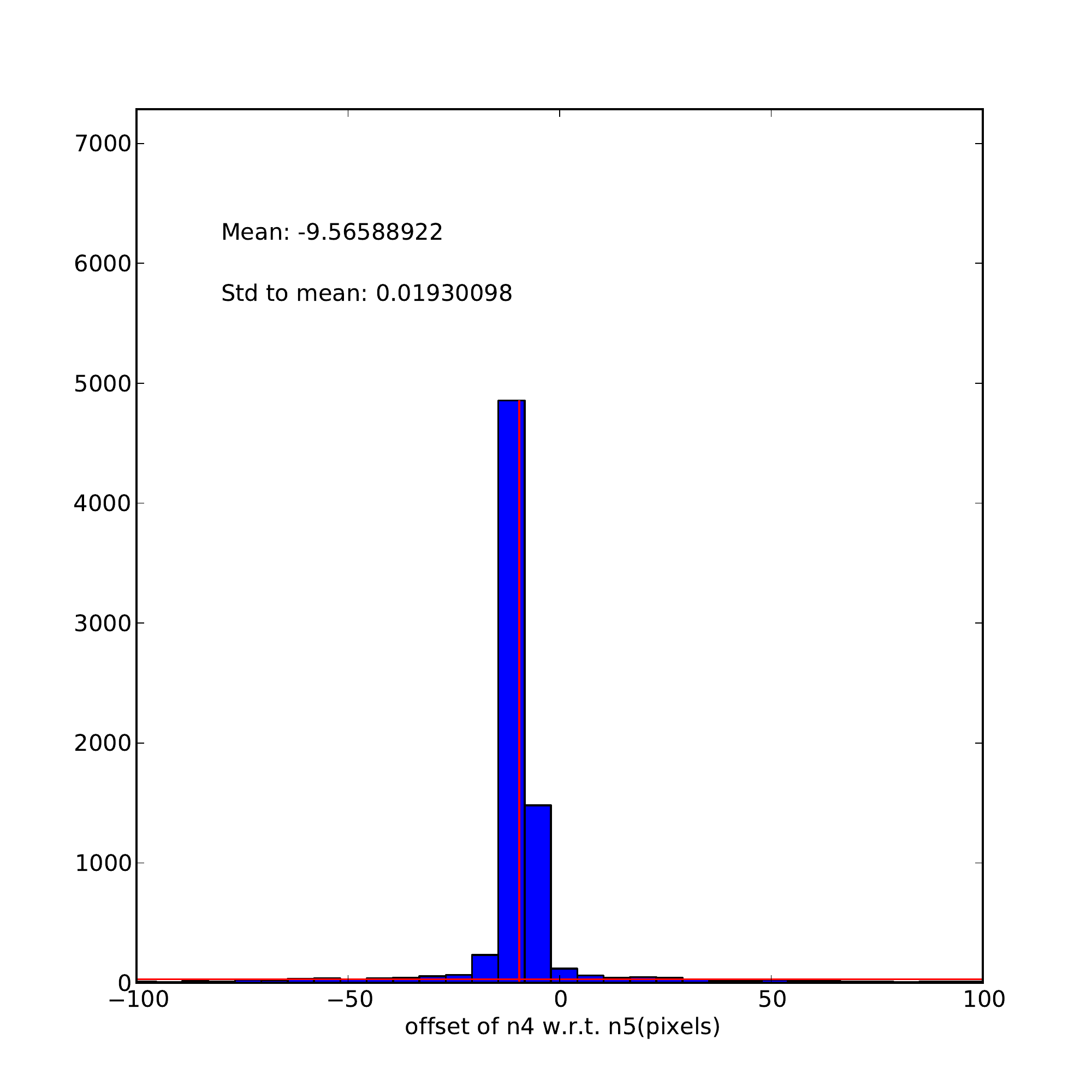}\\
    (a) & (b)\\
   \includegraphics[width=7cm,height=5cm]{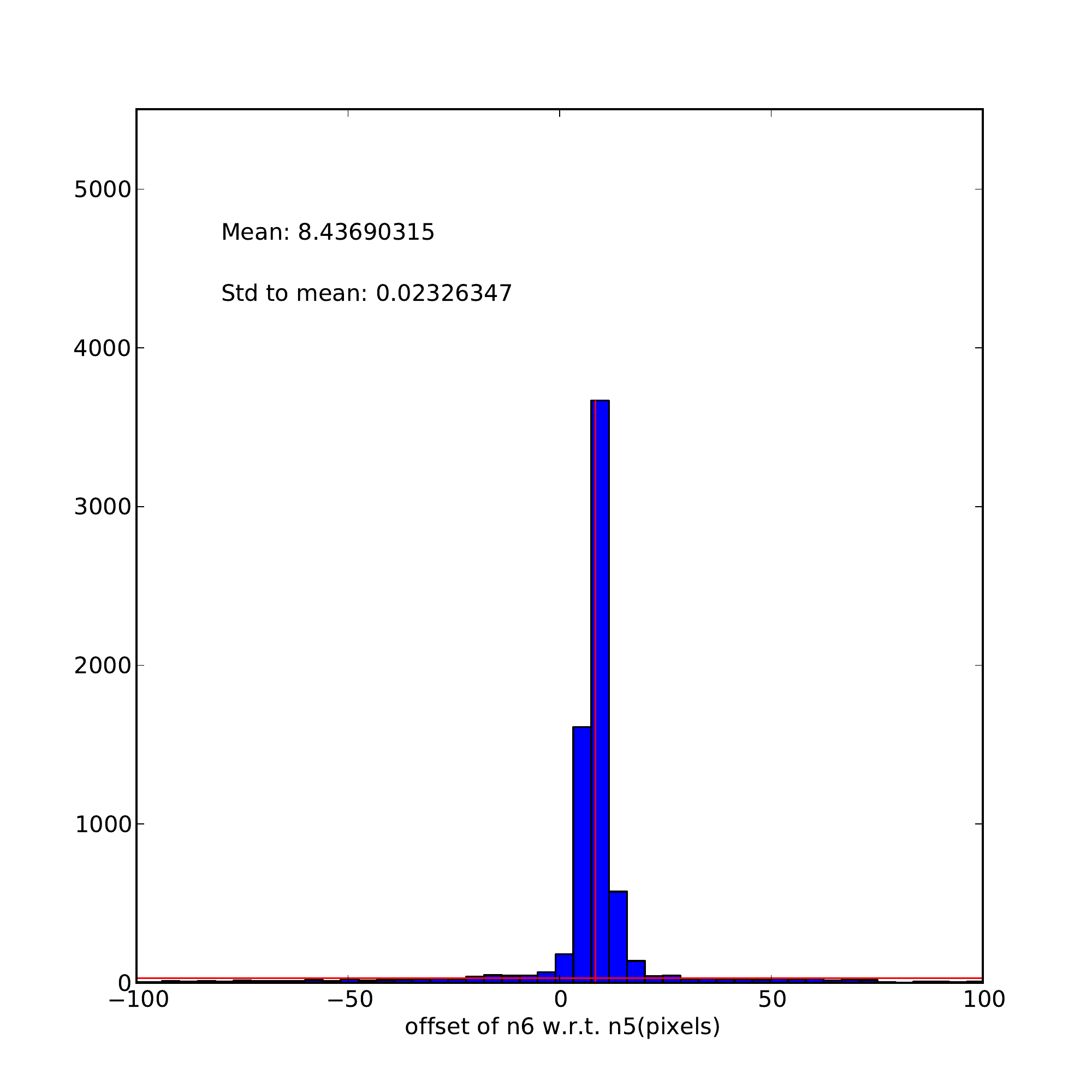} & \includegraphics[width=7cm,height=5cm]{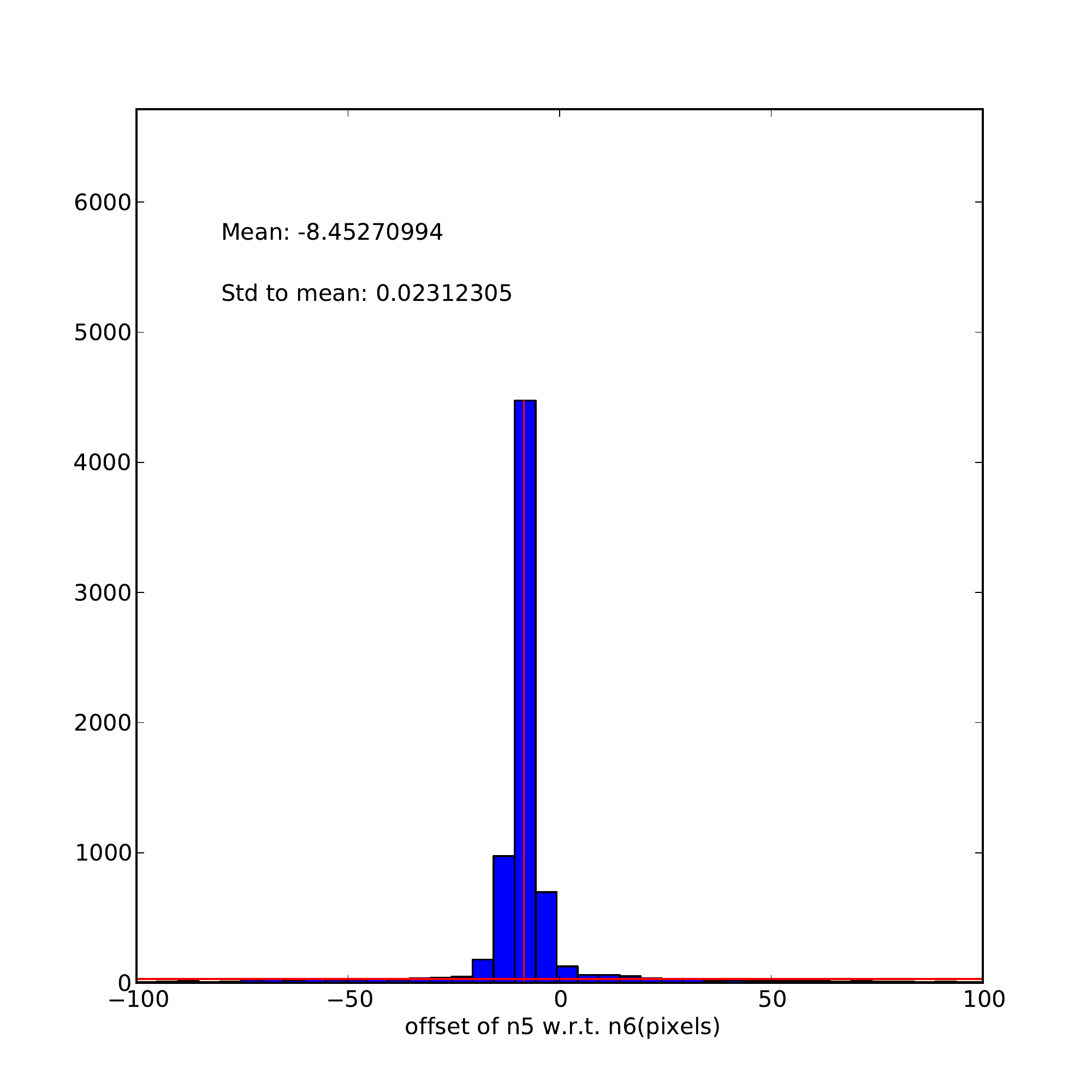}\\
    (c) & (d)\\
   \includegraphics[width=7cm,height=5cm]{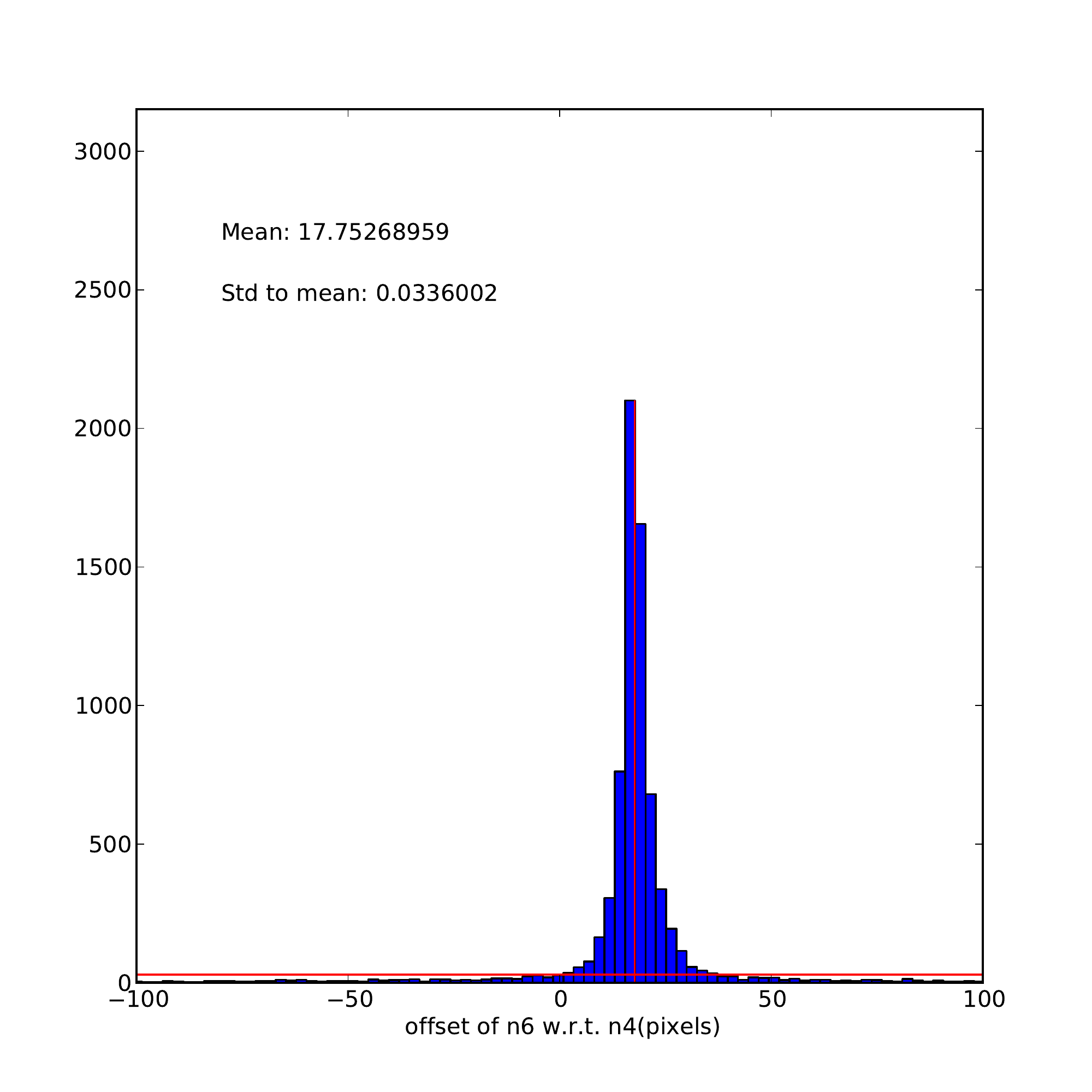} & \includegraphics[width=7cm,height=5cm]{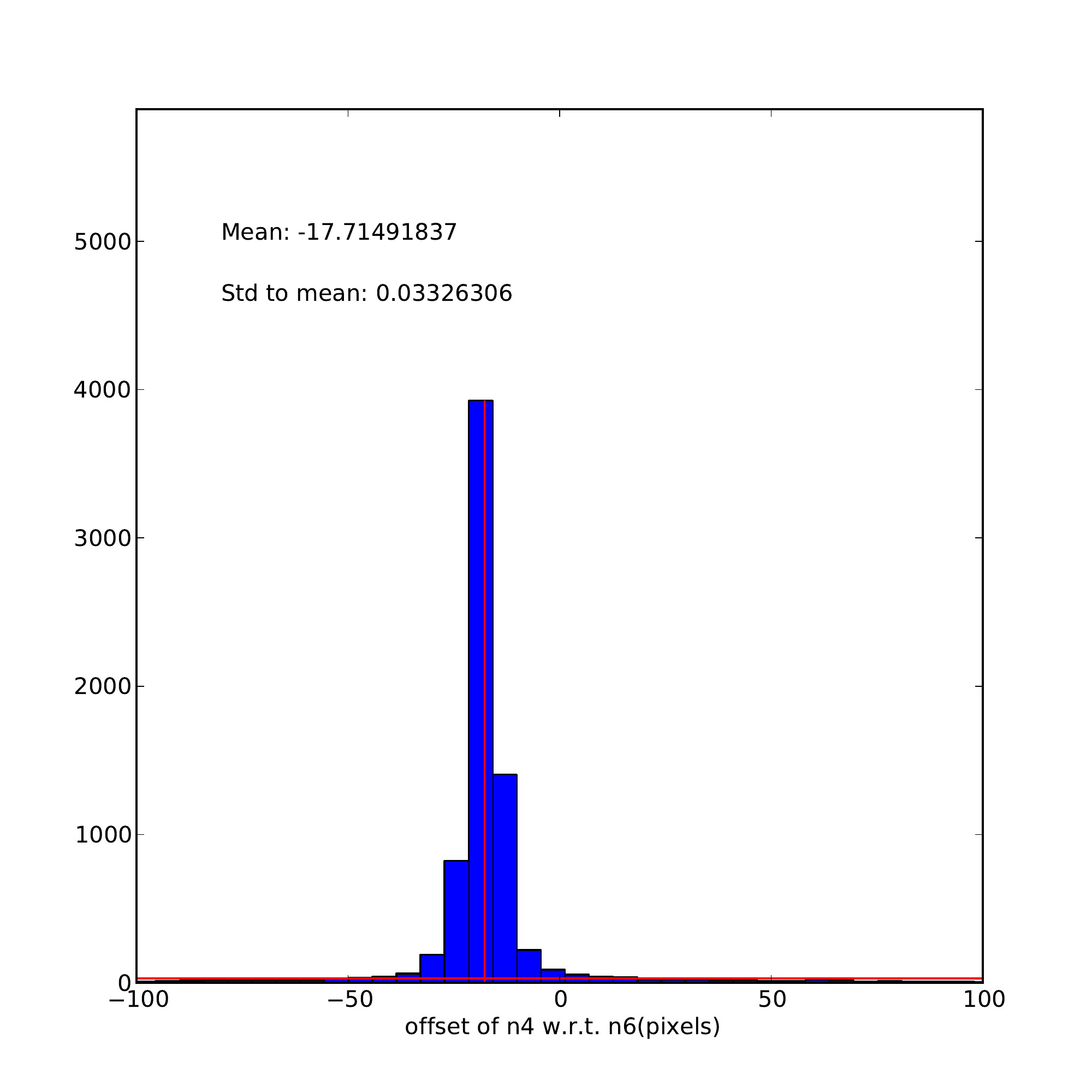}\\
    (e) & (f)  
    \end{tabular}
   \end{center}
   \caption %>>>> use \label inside caption to get Fig. number with \ref{}
   { \label{fig:offset} Offsets among three CCDs. The vertical red line indicate the position of mean while the horizontal red line is a cut we applied to remove those low frequency noisy measurements. This cut only slightly affect the measured quantities since we only remove those very low frequency cases.  
}
   \end{figure}

On our current focal plane, there are 26 CCDs are mounted. We will refter them by their positions in the following. We move the ``star'' projector to each of them and take image. Then, we extract the ``star'' positions and calculate the offsets among the three CCDs. To reduce the bias in the measurements, we measure the offset between two CCDs by turns. That is, we measure the offset of CCD (a) w.r.t to CCD (b) as offset (ab) and then we measure the offset of CCD (b) w.r.t CCD (a) as offset (ba). These two offset should be the same but with opposite sign. We choose the mean of the two offsets (ba and ab) as the final offset between the two CCD. Then, we choose a CCD as the zero point and measure the offset of other CCDs with respect to it. In our case, we choose CCD N4 as our zero point. We list three of such measurements in Figure~\ref{fig:offset} and Table~\ref{table:offset}. 

\begin{table}
   \begin{center}
   \begin{tabular}{|c|c|c|c|}
   \hline
   CCD positions &4N and 5N & 5N and 6N & 4N and 6N \\\hline
   Offset (pixel)& 9.5758 $\pm$ 0.0191 & 8.4448 $\pm$ 0.0232 & 17.7338 $\pm$ 0.0334\\\hline
    \end{tabular}
   \end{center}
   \caption[tab] %>>>> use \label inside caption to get Fig. number with \ref{}
   { \label{table:offset} 
The relative offset among the CCD 4N, 5N and 6N. Looking at the offset, one can see that there is a global tilt between the XY stage plane and the focal plane.}
   \end{table}

\begin{figure}
   \begin{center}
   \begin{tabular}{c c}
   \includegraphics[width=6cm,height=5cm]{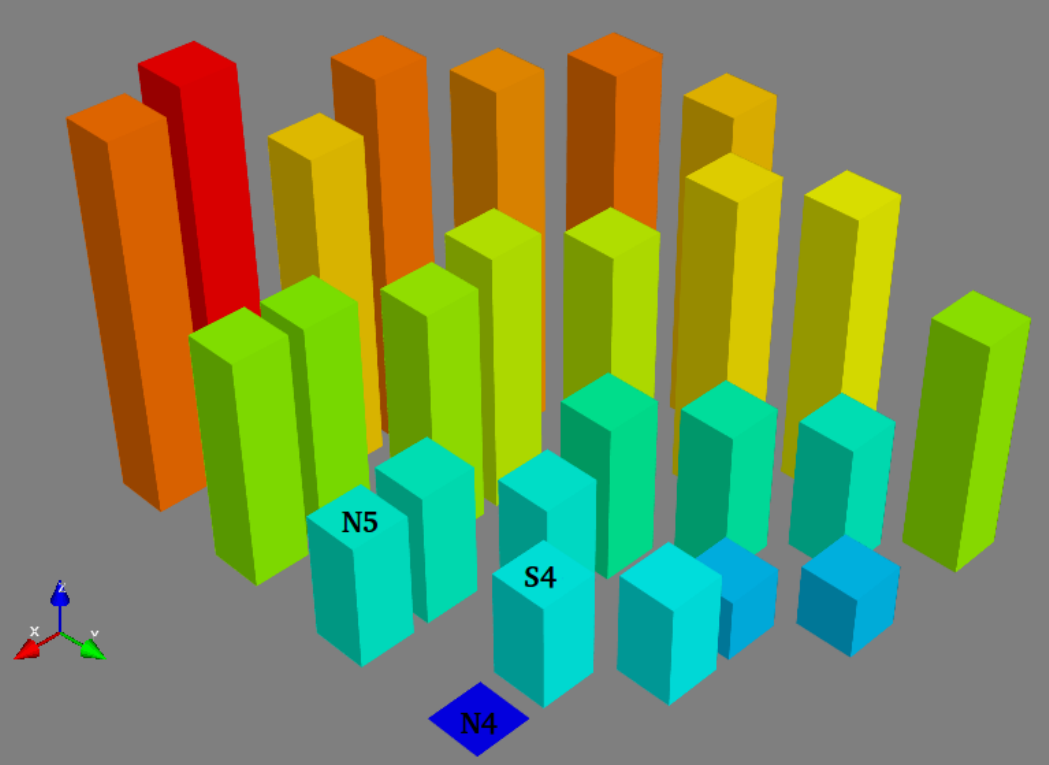} & \includegraphics[width=6cm, height=5cm]{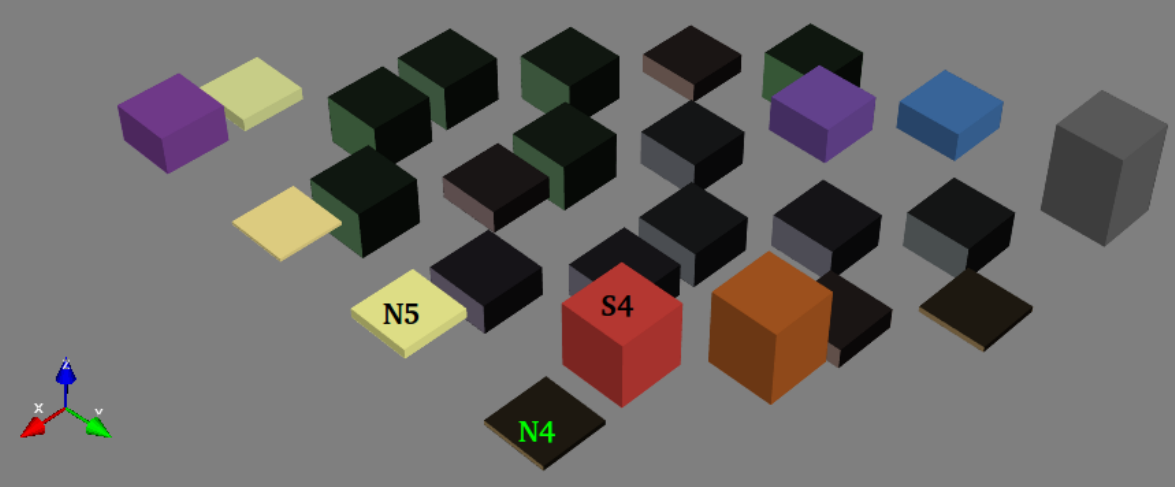}\\
    \end{tabular}
   \end{center}
   \caption %>>>> use \label inside caption to get Fig. number with \ref{}
   { \label{fig:zoffset} The left panel illustrates the offset of CCDs along the optical axis (z direction) before subtracting the best fit plane. The right panel shows the relative offset of each CCD to the best fit plane.   
}
   \end{figure}

\begin{figure}
   \begin{center}
   \includegraphics[width=13cm]{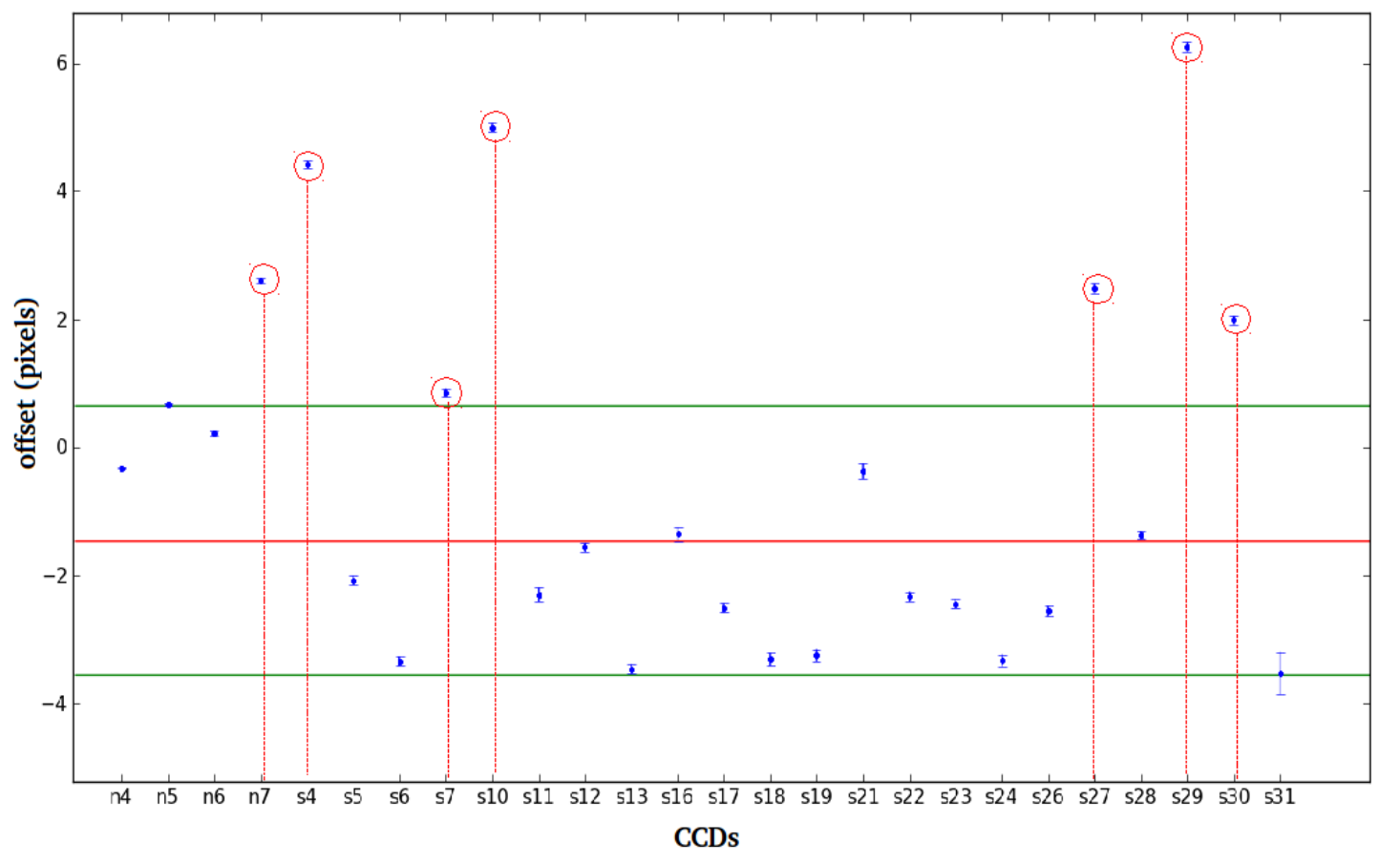}
   \end{center}
   \caption %>>>> use \label inside caption to get Fig. number with \ref{}
   { \label{fig:inspec} The y axis show the relative offsets of each CCD from the best fit plane. The 60-micron envelope is drawn around the median of the relative offsets, indicated by the two green horizontal lines. By this, we have 7 CCD falling out of the envelope.  
}
   \end{figure}

From the results in Table~\ref{table:offset}, one can see a systematic trend, which indicates a global tilt between the XY stage plane and focal plane. Assume the focal plane can be parameterised as $Z = A X + B Y + C$, where X and Y are the coordinate of the centers of the CCDs. After supplying the measurement data, we get a best fit plane with $Z = -0.00149 X -0.00202 Y + 0.3212$. This will correct out the global tilt between focal plane and XY stage plane. Then, each CCD's offset can be measured with respect to this best fit plane. In figure~\ref{fig:zoffset}, we show the 3D illustrations of these offsets w.r.t. zero point (we hold CCD 4N as zero point) and best fit plane. In figure~\ref{fig:inspec}, we show the offset of each CCD w.r.t. the best fit plane and 60-micron envelope around the median of these offsets. We identified 7 CCDs falling outside of the specification, which will be adjusted in our next CCD adjustment operation.

How do we interpret these results more properly? Especially given the fact that each CCD may have intrinsic tilt and we might not apply Equation~\ref{equation:notilt} directly in a strict sense. What we measured are essentially the variation of certain separations between every pair of dots when imaged with different CCDs. On each image, we use many separations to check their corresponding variations by cross the pattern to another image from a different CCD. If we assume that the variations are due to the offset of the CCDs as depicted in Figure~\ref{fig:theory} (a), then the numbers we reported will be the effective offset among CCDs. 

\begin{figure}
   \begin{center}
   \begin{tabular}{c}
   \includegraphics[width=5cm]{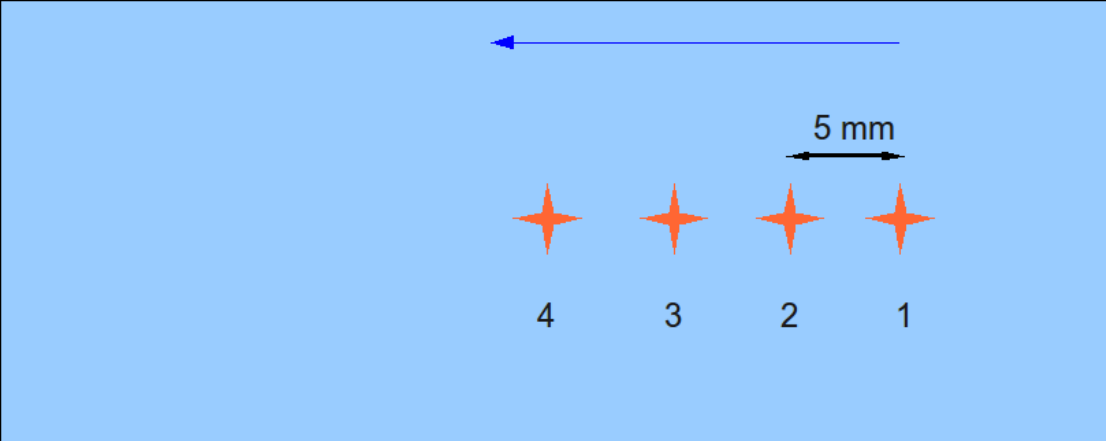}
   \end{tabular}
    \end{center}
   \caption[star moving] 
%>>>> use \label inside caption to get Fig. number with \ref{}
   { \label{fig:star_tilt} Illustration about how we measure the tilt of a single CCD. We move the XY stage from right to left at a increment of 5 mm for each move and take an image at each position. 
}
   \end{figure}

\begin{figure}
   \begin{center}
   \begin{tabular}{c c c}
   \includegraphics[width=5cm]{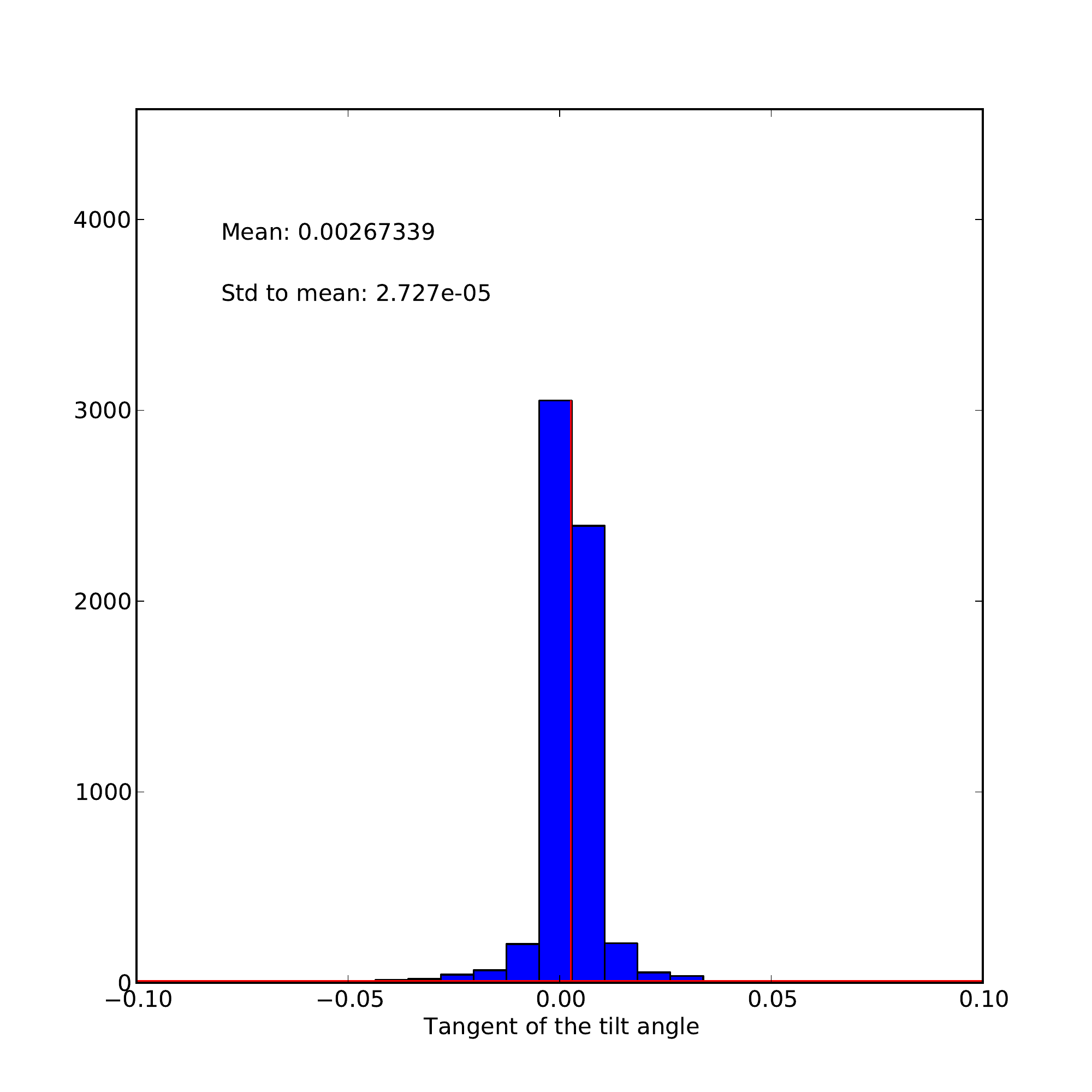}&\includegraphics[width=5cm]{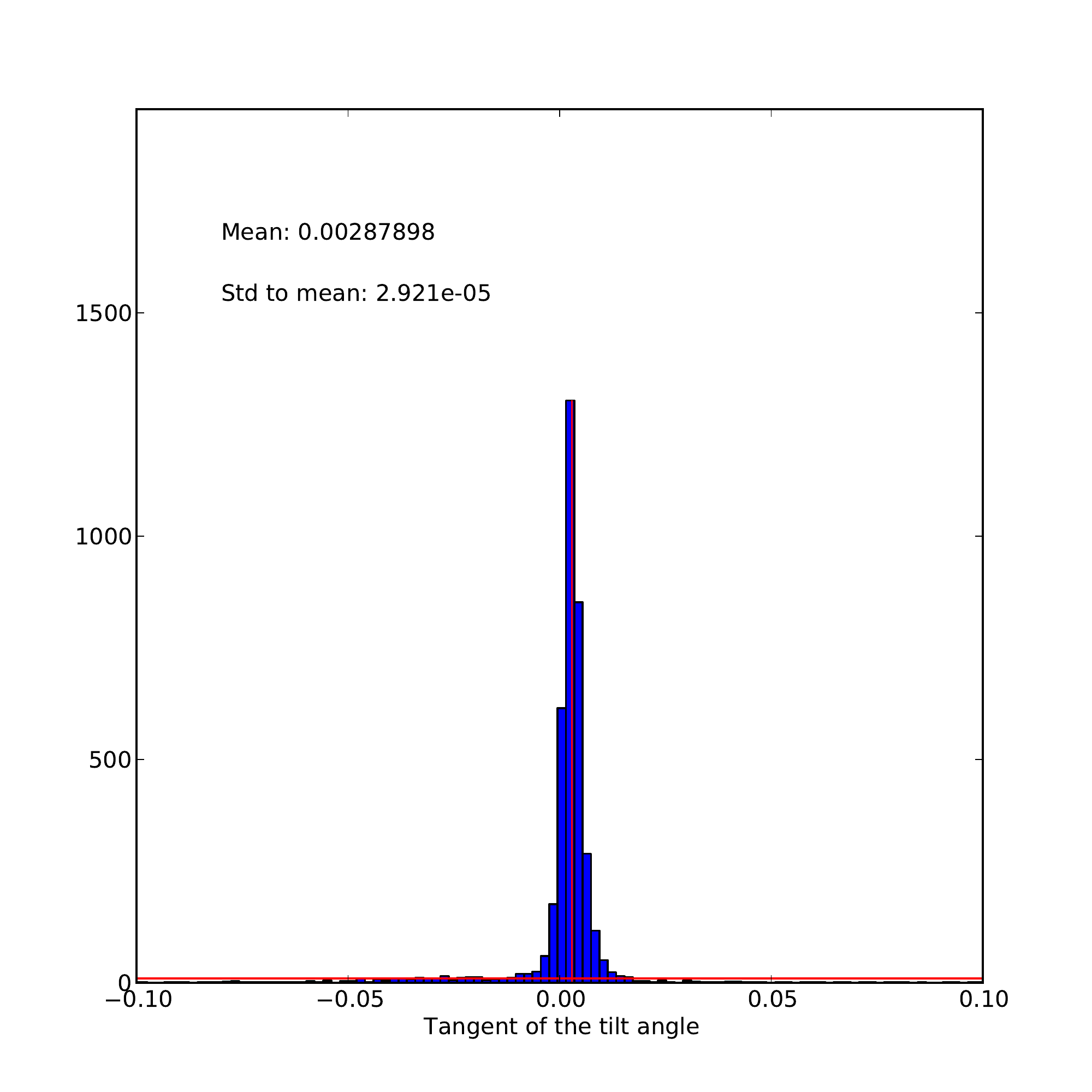}&\includegraphics[width=5cm]{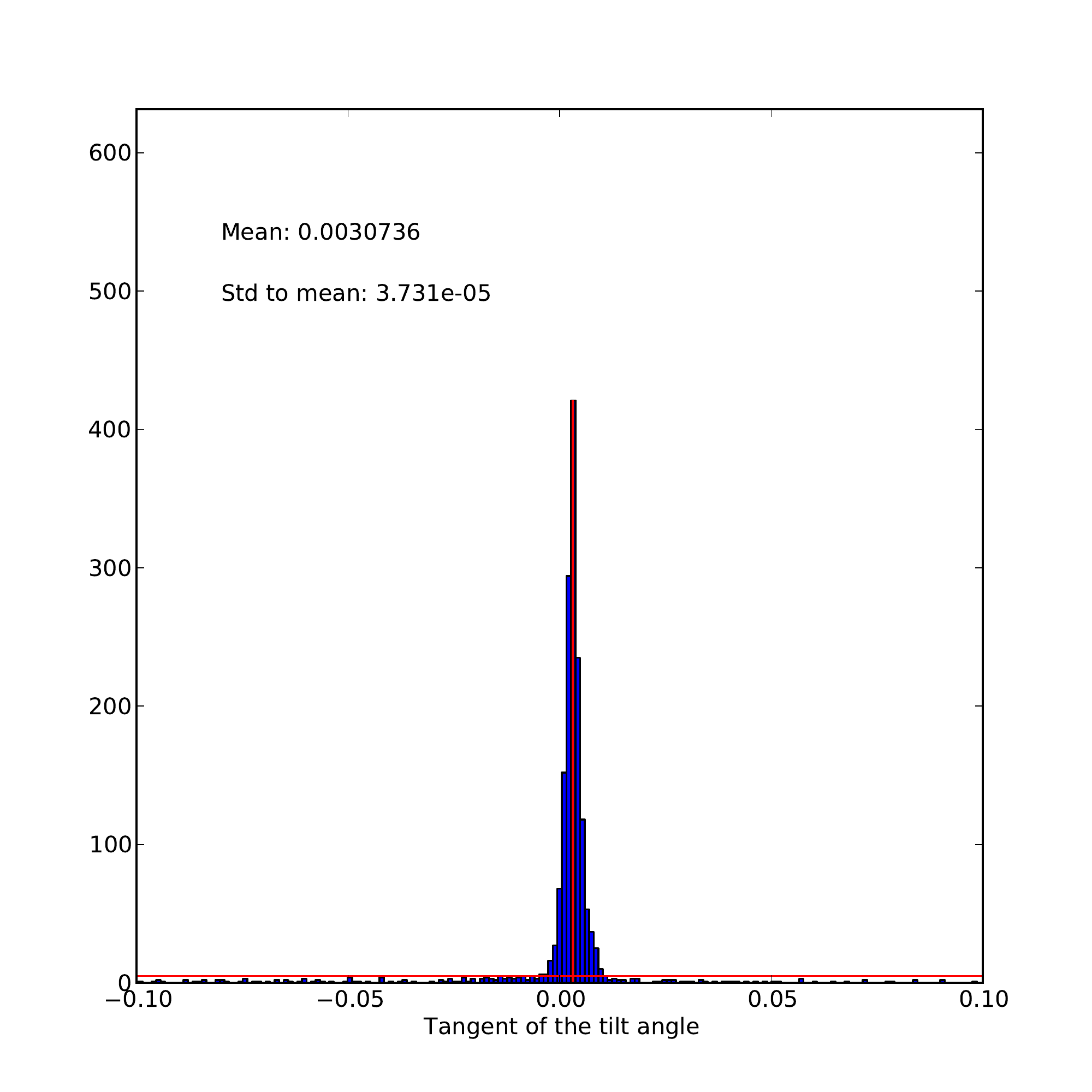}\\
   (a) & (b) & (c)
    \end{tabular}
   \end{center}
   \caption[xystage] 
%>>>> use \label inside caption to get Fig. number with \ref{}
   { \label{fig:tilt}The tangents of a single CCD tilt angle. (a) is the measurement between ``star'' 1 and 2; (b) is the measurement between ``star'' 1 and 3; (c) is the measurement between ``star'' 1 and 4. We choose the mean of this three measurements as the tangent of the CCD tilt angle.
}
   \end{figure}

The previous offset analysis is only for different CCDs. However, we can also look at the offset at several different spots on a single CCD. This will be the case as shown in Figure~\ref{fig:theory} (b). By this, we can measure the tilt of a single CCD. To do this, we took 4 images for CCD 17S as shown in Figure~\ref{fig:star_tilt}. After each image, we move the ``star'' projectors 5 mm left (along the y direction) and then take the next image. Then, we use Equation (2) to calculate the tangent of the tilt angle and the results are shown in Figure~\ref{fig:tilt}. From previous result, we know there is a global tilt between focal plane and XY stage plane. The tangent of this global tilt along y direction is 0.00202, the B parameter of the best fit plane but with opposite sign. So, subtracting the global tilt, the CCD 17S has a intrinsic tilt along y direction of ~0.0008 (the tangent), which are still within the specification when measuring from the center of the CCD.

\section{Discussion}

In this paper, we present two image based methods to measure the flatness of the DECam focal plane. Both simulation and real measurement show the effectiveness of the methods. These measurements are complimentary to other optics based methods and directly measure the image deformation across the focal plane. They are very useful for calibrating the flatness of the focal plane for large mosaic CCD camera. 

\acknowledgments 
Funding for the DES Projects has been provided by the U.S. Department of Energy, the U.S. National Science Foundation, the Ministry of Science and Education of Spain, the Science and Technology Facilities Council of the United Kingdom, the Higher Education Funding Council for England, the National Center for Supercomputing Applications at the University of Illinois at Urbana-Champaign, the Kavli Institute of Cosmological Physics at the University of Chicago, Financiadora de Estudos e Projetos, Fundação Carlos Chagas Filho de Amparo à Pesquisa do Estado do Rio de Janeiro, Conselho Nacional de Desenvolvimento Científico e Tecnológico and the Ministério da Ciência e Tecnologia, the German Research Foundation-sponsored cluster of excellence “Origin and Structure of the Universe” and the Collaborating Institutions in the Dark Energy Survey. 

The Collaborating Institutions are Argonne National Laboratories, the University of California at Santa Cruz, the University of Cambridge, Centro de Investigaciones Energeticas, Medioambientales y Tecnologicas-Madrid, the University of Chicago, University College London, DES-Brazil, Fermilab, the University of Edinburgh, the University of Illinois at Urbana-Champaign, the Institut de Ciencies de l'Espai (IEEC/CSIC), the Institut de Fisica d'Altes Energies, the Lawrence Berkeley National Laboratory, the Ludwig-Maximilians Universität, the University of Michigan, the National Optical Astronomy Observatory, the University of Nottingham, the Ohio State University, the University of Pennsylvania, the University of Portsmouth,  SLAC, Stanford University, and the University of Sussex.

%%% References %%%%%

\bibliography{spieHaoRef}   %>>>> bibliography data in report.bib
\bibliographystyle{spiebib}   %>>>> makes bibtex use spiebib.bst

\end{document}